\documentclass[11pt]{article}

\usepackage{graphicx}
\usepackage{a4wide}

\newcommand{\nit}{\noindent}

\newcommand{\np}{\newpage}
\newcommand{\dsp}{\displaystyle}
\newcommand{\vs}[1]{\vspace{#1 ex}}
\newcommand{\hs}[1]{\hspace{#1 em}}
\newcommand{\bfr}{\begin{flushright}}
\newcommand{\efr}{\end{flushright}}
\newcommand{\bc}{\begin{center}}
\newcommand{\ec}{\end{center}}
\newcommand{\ben}{\begin{enumerate}}
\newcommand{\een}{\end{enumerate}}

\newcommand{\be}{\begin{equation}}
\newcommand{\ee}{\end{equation}}
\newcommand{\ba}{\begin{array}}
\newcommand{\ea}{\end{array}}
\newcommand{\ct}{\cite}
\newcommand{\bit}{\bibitem}
\newcommand{\dd}[2]{\frac{\partial{#1}}{\partial{#2}}}
\newcommand{\ag}{\alpha}
\newcommand{\bg}{\beta}
\newcommand{\gam}{\gamma}
\newcommand{\del}{\delta}

\newcommand{\ve}{\varepsilon}

\newcommand{\thg}{\theta}
\newcommand{\kg}{\kappa}
\newcommand{\lb}{\lambda}
\newcommand{\sg}{\sigma}
\newcommand{\rg}{\rho}

\newcommand{\vf}{\varphi}
\newcommand{\og}{\omega}
\newcommand{\Gam}{\Gamma}
\newcommand{\Del}{\Delta}

\newcommand{\Sg}{\Sigma}

\newcommand{\brg}{\bar{g}}

\newcommand{\brx}{\bar{x}}

\newcommand{\brR}{\bar{R}}

\newcommand{\brGam}{\bar{\Gam}}

\newcommand{\lh}{\left(}
\newcommand{\rh}{\right)}
\newcommand{\ld}{\left.}

\newcommand{\der}{\partial}

\begin{document}

\pagestyle{empty} 

\hfill NIKHEF/2010-042
\vs{7}

\bc
{\Large{\bf Epicycles and Poincar\'{e}  Resonances}} \\
\vs{2}

{\Large{\bf in General Relativity}}
\vs{7}

{\large{G.\ Koekoek$^1$}} 
\vs{2}

{\large{Physics Department, Vrije Universiteit Amsterdam}}
\vs{3} 

{{\large J.W.\ van Holten$^2$}}
\vs{2}

{\large{Nikhef, Amsterdam}}
\vs{3}

November 15, 2010
\vs{10}

{\small{\bf Abstract}}
\ec

\nit
{\small
The method of geodesic deviations provides analytic approximations to geodesics in arbitrary
background space-times. As such the method is a useful tool in many practical situations. In this 
note we point out some subtleties in the application of the method related to secular motions, in
first as well as in higher order. In particular we work out the general second-order contribution 
to bound orbits in Schwarzschild space-time and show that it provides very good analytical results 
all the way up to the innermost stable circular orbit. }

\vfill
\footnoterule 
\nit
{\footnotesize{$^1$ e-mail: gkoekoek@nikhef.nl \\
$^2$ e-mail: v.holten@nikhef.nl }}

\np
\pagestyle{plain}
\pagenumbering{arabic}

\section{Geodesic deviations \label{s1}}

According to General Relativity, in a fixed background space-time compact objects (test masses) move
on geodesics. However, except for the simplest cases, their orbits can be calculated from the geodesic 
equation only in certain approximations.  An often-used method is provided by the post-newtonian 
approximation scheme, which starts from the non-relativistic orbit, and then systematically calculates 
special and general relativistic corrections \ct{mtw, hartle, fut-itoh}. 

A different approximation scheme is provided by the method of geodesic deviations \ct{mtw, hartle}. This 
is a manifestly covariant method, which can be extended to include other interactions, e.g.\ in Einstein-Maxwell
theory \ct{bkh, kmmh} and non-abelian backgrounds \ct{bgsw}, or the effects of spin \ct{jwvh,moh}. Moreover, 
the method can be extended to arbitrary precision by taking into account higher-order deviations \ct{khc,ker-col}. 
In the literature cited \ct{bkh}-\ct{ker-col}, the method has been applied e.g.\ to calculate particle orbits in pp-waves, 
and Schwarzschild, Reissner-Nordstr{\o}m and Kerr space-times. 

In this paper we consider the application of the geodesic deviation method to pure gravity, pointing out some 
subtleties which arise already at linear order, but which become of special relevance when extending the 
method to include higher-order corrections. Similar subtleties also arise in newtonian gravity,  and
have been adressed in newtonian perturbation theory already in the 19th century \ct{lind, poinc}. 

The geodesic deviation method starts from a known reference geodesic, and then computes neighboring
geodesics by determining the space-time vector connecting the points of the reference orbit with points on 
the unknown geodesic. Let $x^{\mu}[\tau; \sg]$ represent a continuous family of geodesics, the proper time
$\tau$ acting as affine parameter, and $\sg$ labeling the geodesics in the family. Let $\brx^{\mu}(\tau) 
= x^{\mu}[\tau; 0]$ be a known geodesic:
\be
\frac{D^2 \brx^{\mu}}{D\tau^2} = \frac{d^2 \brx^{\mu}}{d\tau^2} + \brGam_{\lb\nu}^{\;\;\;\mu} \frac{d\brx^{\lb}}{d\tau}
 \frac{d\brx^{\nu}}{d\tau} = 0,
\label{1.1}
\ee
where $\brGam = \Gam[\bar{x}(\tau)]$ represents the connection evaluated on the  geodesic $\brx[\tau]$. Neighboring 
geodesics are then found from the expansion
\be
x^{\mu}[\tau; \sg] = \brx^{\mu}(\tau) + \sg n^{\mu}(\tau) + \frac{1}{2}\, \sg^2 m^{\mu}(\tau) + ...
 =  \brx^{\mu} + \sg n^{\mu} + \frac{1}{2}\, \sg^2 \lh k^{\mu} - \brGam_{\lb\nu}^{\;\;\;\mu} n^{\lb} n^{\nu} \rh + ...
\label{1.2}
\ee
In this expansion the vectors $n^{\mu}$ and $k^{\mu}$ are defined covariantly as 
\be
n^{\mu} = \ld \dd{x^{\mu}}{\sg} \right|_{\sg = 0}, \hs{2} 
k^{\mu} = \ld \frac{Dn^{\mu}}{D\sg} \right|_{\sg = 0} = \ld \dd{n^{\mu}}{\sg} \right|_{\sg = 0} 
 + \brGam_{\lb\nu}^{\;\;\;\mu}  n^{\lb} n^{\nu}.
\label{1.3}
\ee
The scale can be set, for example, by defining $\sg$ to represent the proper distance along the curve 
$x^{\mu}[0; \sg]$:
\be
d\sg^2 = \ld g_{\mu\nu}\, dx^{\mu} dx^{\nu} \right|_{\tau = 0} \hs{1} \Rightarrow \hs{1}
\ld \bar{g}_{\mu\nu} n^{\mu} n^{\nu} \right|_{\tau = 0} = 1,
\label{1.3.1}
\ee
where, as for the connection,  $\bar{g}_{\mu\nu} = g_{\mu\nu}[\bar{x}(\tau)]$ is the metric evaluated on the reference 
geodesic $\bar{x}^{\mu}(\tau)$. Of course, as the distance between geodesics varies the normalization of $n$ is in 
general not preserved as a function of proper time. Specifically, the change of the geodesic deviation vectors 
$(n, k, ... )$ is determined by the geodesic deviation equations
\be
\ba{l} 
\dsp{ \frac{D^2 n^{\mu}}{D\tau^2} - \brR_{\lb\nu\kg}^{\;\;\;\;\;\mu} u^{\kg} u^{\lb} n^{\nu} = 0, }\\
 \\
\dsp{ \frac{D^2 k^{\mu}}{D\tau^2} - \brR_{\lb\nu\kg}^{\;\;\;\;\;\mu} u^{\kg} u^{\lb} k^{\nu} = 
 \brR_{\lb\nu\kg\; ;\rg}^{\;\;\;\;\;\mu} \lh u^{\kg} u^{\lb} n^{\nu} n^{\rg} - u^{\nu} u^{\rg} n^{\kg} n^{\lb} \rh
 + 4 \brR_{\lb\nu\kg}^{\;\;\;\;\;\mu} u^{\lb} n^{\nu} \frac{Dn^{\kg}}{D\tau}, }
\ea
\label{1.4}
\ee
and higher-order generalizations \ct{khc}, where $u^{\mu}$ is the four-velocity along the reference geodesic:
\be
u^{\mu} = \frac{d\brx^{\mu}}{d\tau} = \ld \dd{x^{\mu}}{\tau} \right|_{\sg = 0}.
\label{1.5}
\ee
By construction, these equations are successively linear in the perturbation vectors $(n,k,...)$; however, 
whereas the equation for the first-order perturbation is a homogeneous linear differential equation, the 
equation for the higher-order corrections are inhomogeneous linear differential equations. The inhomogeneous 
terms are polynomial in the lower-order perturbations, such that e.g.\ the second-order perturbation $k$ is 
determined by terms quadratic in the first-order perturbation $n$. In these inhomogeneous quadratic terms 
$n$ is to be considered as a known vector, having been solved from the homogeneous first-order equation. 

Another obvious property of these equations is, that the linear terms on the left-hand side are all of the same 
form \ct{khc}. This implies that the general solution of the equation for the second- and higher-order perturbations 
is some particular solution plus an arbitrary solution of the homogeneous equation, i.e.\ a first-order solution. 

Considering the first-order equation, we observe that its general form is that of a parametric oscillator, with
a proper-time dependent driving force linear in the amplitude $n$, the full proper-time dependence being 
determined by the curvature tensor and the four-velocity on the reference geodesic. The full power of the 
geodesic deviation method is developed if the manifold considered admits a reference geodesic for which 
the parametric force equation is solvable. Such solutions can be found in many important cases, such as the 
Schwarzschild, Reissner-Nordstr{\o}m and Kerr solutions in four dimensional space-time. Below we consider 
in particular the Schwarzschild geometry, but the methodological observations hold quite generally.

\section{First-order perturbations \label{s2}} 

The linear first-order equation is the key to all higher-order ones, as it provides both inhomogenous
terms for, and the general homogeneous solutions of, the higher-order equations. In spaces with symmetries 
there are some obvious solutions. Indeed, if $\xi^{\mu}(x)$ is a Killing vector field:
\be
\xi_{\nu;\mu} + \xi_{\mu;\nu} = 0,
\label{2.1}
\ee
it generates an isometry of the metric and it follows quite simply that it also satisfies the geodesic deviation
equation:
\be
\frac{D^2 \xi^{\mu}}{D\tau^2} = R_{\lb\nu\kg}^{\;\;\;\;\;\mu} u^{\lb} u^{\kg} \xi^{\nu},
\label{2.2}
\ee
along {\em any} geodesic. For example, in exterior Schwarzschild space-time, which is static and spherically
symmetric, any geodesic is turned into another geodesic by a time translation or a rotation. By using this 
property one can in fact reduce the class of geodesics to be investigated by simply modding out part of the rotations, 
and considering only geodesics in the equatorial plane, as in the standard textbook treatment \ct{mtw,hartle,chandra}. 
When discussing the geodesics of Schwarzschild space-time below we do the same. 

Another interesting situation occurs, if geodesics are simultaneously lines of Killing flow, i.e.\ if a Killing vector is
transported parallel to itself along a geodesic. Then the tangent vector (four-velocity) along such a geodesic is 
itself a Killing vector $\xi^{\mu}$, and its norm is constant along the geodesic
\be
\der_{\mu} \xi^2 = 2\, \xi^{\nu} \xi_{\nu;\mu} = -2\, \xi^{\nu} \xi_{\mu;\nu} = 0.
\label{2.3.0}
\ee
An example is provided by circular equatorial orbits in Schwarzschild geometry (our conventions are summarized
in appendix \ref{saa}), which are characterized by a four-velocity  
\be
u^{\mu} = - \ve_0\, \xi_t^{\mu} + \ell_0\, \xi_{\vf}^{\mu}.
\label{2.3}
\ee
Here $(\xi_t, \xi_{\vf})$ are the Killing vectors generating time translations and axial rotations, and 
$(\ve, \ell)$ are the corresponding constants of motion defined in appendix \ref{saa}, eq.\ (\ref{aa.2}), 
which on circular orbits take the values 
\be
\ve_0^2 = \lh 1 - \frac{2M}{R} \rh^2  \lh 1 - \frac{3M}{R} \rh^{-1}, \hs{1}
\ell_0^2 = MR \lh 1 - \frac{3M}{R} \rh^{-1}.
\label{2.3.1}
\ee
Explicitly, the Killing vectors for time translations and axial rotations are represented by the differential operators
\be
\xi_t = \xi_t^{\mu} \der_{\mu} = g^{tt} \der_t , \hs{2} \xi_{\vf} = \xi_{\vf}^{\mu} \der_{\mu} = g^{\vf\vf} \der_{\vf}.
\label{2.5}
\ee
These Killing vectors are orthogonal: $\xi_t \cdot \xi_{\vf} = 0$, and eqs.\ (\ref{2.3.1}) imply that the linear 
combination (\ref{2.3}) is time-like and normalized:
\be
u^2 = \ve_0^2\, \xi_t^2 + \ell_0^2\, \xi_{\vf}^2 = -1.
\label{2.6}
\ee
Returning to the general discussion, if the tangent vector $u^{\mu}$ generates an isometry, the metric and 
connection are constant along the geodesic. With such a geodesic as a reference, the geodesic deviation 
equations then reduce to ordinary linear second-order differential equations with {\em constant} co-efficients. 
In particular, for bound orbits these equations have standard solutions with real eigenfrequencies \ct{bkh,khc}. 

To find explicit solutions, it is convenient to write out the first-order equation in its non-manifestly covariant 
form
\be
\frac{d^2 n^{\mu}}{d\tau^2} + 2u^{\lb} \brGam_{\lb\nu}^{\;\;\;\mu} \frac{dn^{\nu}}{d\tau} 
 + u^{\kg} u^{\lb} \der_{\nu} \brGam_{\kg\lb}^{\;\;\;\mu}\, n^{\nu} = 0.
\label{2.0}
\ee
As for the special reference orbits the coefficients are constant, the generic solution for bound orbits will be periodic: 
\be
n_{per}^{\mu} = n_c^{\mu} \cos \og \tau + n_s^{\mu} \sin \og \tau,
\label{2.0.0}
\ee
with constant amplitudes $(n_c, n_s)$. The eigenfrequencies $\og$ are found by diagonalizing the
characteristic matrix for the differential equations (\ref{2.0}), defined by the set of linear equations
\be
\ba{l}
\dsp{ - \og^2 n_c^{\mu} + 2 \og u^{\lb} \brGam_{\lb\nu}^{\;\;\;\mu} n_s^{\nu} +
 u^{\kg} u^{\lb} \der_{\nu} \brGam_{\kg\lb}^{\;\;\;\mu}\, n_c^{\nu} = 0, }\\ 
  \\
\dsp{ - \og^2 n_s^{\mu} - 2 \og u^{\lb} \brGam_{\lb\nu}^{\;\;\;\mu} n_c^{\nu} +
 u^{\kg} u^{\lb} \der_{\nu} \brGam_{\kg\lb}^{\;\;\;\mu}\, n_s^{\nu} = 0. }
\ea
\label{2.7}
\ee
It follows, that in $n$ space-time dimensions the characteristic equation in general has $2n$ roots. 
However, even if the reference orbit is strictly bound (i.e., it is enclosed in a finite region of space) 
and the eigenfrequencies are guaranteed to be real, there can still be zero-modes. These zero modes
take the form of polynomials in $\tau$, rather than periodic functions of the type (\ref{2.0.0}). Being 
non-periodic, they describe secular motions. The importance of, and how to deal with, such zero modes 
is addressed in the following. 

\section{First-order geodesics in Schwarzschild space-time \label{s2.1}}

Before continuing the general discussion, we analyze in more detail the example of orbits near a 
circular orbit in exterior Schwarzschild space-time. We have already seen that for such orbits we 
can reduce the problem from a four-dimensional one to a three-dimensional one by restriction to 
the equatorial plane. For motions in the equatorial plane, eq.\ (\ref{2.0}) takes the form
\be
\lh \ba{ccc} 
\frac{d^2}{d\tau^2} & \ag \frac{d}{d\tau} & 0 \\
 & & \\
 \bg \frac{d}{d\tau} & \frac{d^2}{d\tau^2} - \kg & - \gam \frac{d}{d\tau} \\
 & & \\
 0 & \eta \frac{d}{d\tau} & \frac{d^2}{d\tau^2} \ea \rh \lh \ba{c} n^t \\ \\ n^r \\ \\ n^{\vf} \ea \rh = 0,
\label{2.0.1}
\ee
where
\be
\ba{lll}
\ag = \frac{2M}{R^2 \lh 1 - \frac{2M}{R} \rh} \frac{1}{\sqrt{ 1 - \frac{3M}{R} }}, & 
\bg = \frac{2M}{R^2} \frac{1 - \frac{2M}{R}}{\sqrt{ 1 - \frac{3M}{R} }}, & 
\gam = 2 \sqrt{\frac{M}{R}} \frac{1 - \frac{2M}{R}}{\sqrt{ 1 - \frac{3M}{R} }}, \\
 & & \\
\eta = \frac{2}{R^2} \sqrt{\frac{M}{R}} \frac{1}{\sqrt{ 1 - \frac{3M}{R} }}, & 
\kg = \frac{3M}{R^3} \frac{1 - \frac{2M}{R}}{1 - \frac{3M}{R}}, & 
\ea
\label{2.0.2}
\ee
Then the linear equations (\ref{2.7}) form a six-dimensional system, and in terms of the eigenfrequencies
$\og$ the characteristic equation takes the form \ct{khc}
 \be
 \og^4 \left[ \og^2 - \eta \gam + \ag \bg + \kg \right] = 0,  
 \label{2.8}
 \ee
 showing explicitly the existence of zero-modes even upon restriction of the problem
 to the equatorial plane. 
 
Choosing initial conditions $t(0) = \vf(0) = 0$, the periodic solutions (\ref{2.0.0}) satisfy
\be
\og n^t_s = - \ag n^r_c, \hs{2} \og n^{\vf}_s = -\eta n^r_c, \hs{2} 
\og^2 = \eta \gam - \ag \bg - \kg = \frac{M}{R^3} \frac{1 - \frac{6M}{R}}{1 - \frac{3M}{R}},
\label{2.9}
\ee
whilst $n^t_c = n^{\vf}_c = n^r_s = 0$. 
As the period of the geodesic deviation (\ref{2.0.0}) --which can be interpreted as the relativistic 
generalization of an {\em epicycle}-- differs from that of the circular orbit we started from, the point of 
closest approach (the periastron) shifts during each orbit by a fixed amount. This accounts for the 
well-known precession of the periastron in general relativity \ct{bkh, khc}. 

The zero-modes correspond to secular motions described by linear functions
\be
n_{sec}^{\mu} = v^{\mu} \tau + \Del_n^{\mu}, 
\label{2.10}
\ee
with
\be
v^r = 0, \hs{2}
\kg\, \Del^r_n = \bg v^t - \gam v^{\vf}.
\label{2.12}
\ee
Observe, that a non-zero value for $v^r$ would have been unacceptable, as it would contradict the boundedness
of the orbits described. However, non-zero values for $v^t$ and $v^{\vf}$ do not cause such problems and are 
perfectly allowed. In fact, such solutions are required for at least two reasons. First, as observed in \ct{khc}, the
inhomogeneous terms in the higher-order geodesic perturbation equations generate Poincar\'{e} resonances
which have to be removed by such secular terms. This is discussed in detail below. Second, even in the first-order 
approximation the periodic solutions suffer from the problem that the angle and time between periastra can not be 
matched correctly for eccentric orbits; this mismatch accumulates and grows without bound in due course of time, 
unless corrected by the secular terms (\ref{2.10}), (\ref{2.12}).

Whilst a non-zero value of $\Del^r_n$ is required by the position and precession of the periastron and implies
non-zero values for $(v^t, v^{\vf})$, the shifts in the cyclic co-ordinates $\Del_n^t$ and $\Del_n^\vf$ only change 
the origin of time and azimuth angle and are therefore arbitrary. In view of our initial conditions we take 
$\Del_n^t = \Del_n^{\vf} = 0$. As $\Del_n^r$ then is the only remaining relevant component, from here on we will 
for simplicity write $\Del^r_n = \Del_n$. The secular solutions (\ref{2.10}) are further restricted by the normalization 
of the four-velocity. To first order this restriction takes the form
\be 
u_{\mu} \frac{Dn^{\mu}}{D\tau} = 0 \hs{1} \Rightarrow \hs{1}  \ve_0 v^t - \ell_0 v^{\vf} = 0.
\label{2.13}
\ee
Taking into account eq.\ (\ref{2.12}) the solutions for the secular velocity terms are
\be
v^t = \frac{\ag \kg}{\ag \bg - \eta \gam}\, \Del_n, \hs{1} v^{\vf} = \frac{\eta \kg}{\ag \bg - \eta \gam}\, \Del_n.
\label{2.14}
\ee
Furthermore observe, that application of the normalization condition (\ref{1.3.1}) implies
\be
n^r(0) = n_{per}^r(0) + n_{sec}^r(0) = n^r_c + \Del_n = \sqrt{1 - \frac{2M}{R}}.
\label{2.9.1}
\ee
Combining all results, the solutions for the first-order perturbed geodesics describing bound motion become
\be
\ba{l}
\dsp{ t = \frac{\tau}{\sqrt{ 1 - \frac{3M}{R} }} - \frac{\ag}{\og}\, \sg n_c^r \sin \og \tau 
 + \frac{\ag \kg \tau}{\ag \bg - \gam \eta}\, \sg \Del_n, }\\
 \\
\dsp{ r = R + \sg n^r_c \cos \og \tau + \sg \Del_n, }\\
 \\
\dsp{ \vf = \sqrt{\frac{M}{R^3}} \frac{\tau}{\sqrt{ 1 - \frac{3M}{R} }} - \frac{\eta}{\og}\, \sg n_c^r \sin \og \tau 
 + \frac{\eta \kg \tau}{\ag \bg - \gam \eta}\, \sg \Del_n.}
\ea 
\label{2.15}
\label{G26}
\ee
We now relate the parameters $\Del_n$ and $\sg$ to observable quantities. First, note that the
periastra and apastra of the orbit occur at proper times $\tau_n$ such that $\og \tau_n = n \pi$. We take the
even values of $n$ to correspond to closest approach (periastron, {\em pa}) and the odd values of $n$ to 
maximal distance (apastron, {\em aa}). Then we have
\be
r_{pa} = R  + \sg \lh \Del_n + n^r_c \rh, \hs{2} r_{aa} = R + \sg \lh \Del_n - n^r_c \rh. 
\label{2.15.1}
\ee
Inverting these equations we get
\be
\sg n_c^r = \frac{1}{2} \lh r_{pa} - r_{aa} \rh, \hs{2} 
\sg \Del_n = \frac{1}{2} \lh r_{pa} + r_{aa} \rh - R.
\label{2.15.2}
\ee
In view of eq.\ (\ref{2.9.1}) it follows that $\sg$ takes the value
\be
\sg = \frac{r_{pa} - R}{\sqrt{1 - \frac{2M}{R}}},
\label{2.15.3}
\ee
hence the dimensionless parameter $\sg/R$ is related to the eccentricity of the orbit. Starting from these 
expressions we can calculate the energy and angular momentum per unit of mass for these perturbed orbits: 
\be
\ve_n = \ve_0 + \del \ve = \lh 1 - \frac{2M}{r} \rh \frac{d t}{d\tau}, \hs{2}
\ell_n = \ell_0 + \del \ell = r^2\, \frac{d \vf}{d\tau},
\label{2.17}
\ee
to find the first-order change in the constants of motion as compared to the circular orbits:
\be 
\lh 1 - \frac{2M}{R} \rh \frac{\del \ve}{\ve_0} = \frac{M}{R}\, \frac{\del \ell}{\ell_0} 
 = \frac{\sg}{2}\, \og^2 R\, \Del_n.
\label{2.16}
\label{G30}
\ee
It follows directly, that the values of $(\ve, \ell)$ are unchanged if and only if the secular contributions vanish: 
$\Del_n = v^t = v^{\vf} = 0$. However, in general we want to allow changes of these values, as this is required 
to describe also non-circular orbits.

More precisely: up to orientation bound geodesics in Schwarzschild space-time are characterized by the 
two constants of motion $\ve$ and $\ell$. Together these two parameters determine the angle as well as the 
time lapse between successive periastra. However, for circular orbits these parameters are not independent, 
as both are determined completely by the value of the radial co-ordinate $R$. Therefore it is not possible to 
choose a zeroth-order  geodesic which is circular and has arbitrary independent preassigned values of $\ve$ 
and $\ell$. In contrast, at first order it becomes possible to adjust the parameters such that $\ve$ and $\ell$ 
can be given independent values, but only by changing these constants of motion compared to the values 
they have on the circular geodesic. Therefore in general one has to choose a non-zero value for the secular 
contributions from non-vanishing $\Del_n$. In practice it is easiest to fix the periastron parameters (radial 
distance, angle and time lapse between successive periastra) to have preassigned values, relating them 
to $\ve$ and $\ell$ afterwards by the procedure described above.

\section{Second-order perturbations \label{s3}}

The first-order geodesic deviations are solutions of the coupled homogeneous linear differential equations (\ref{2.0})
and combine both periodic and secular terms. In the second-order deviation equations these solutions reappear in 
the inhomogeneous terms. This is evident from the second equation (\ref{1.4}), which can again be cast in a 
non-covariant but more tractable form in terms of $m^{\mu} = k^{\mu} - \brGam_{\lb\nu}^{\;\;\;\mu} n^{\lb} n^{\nu}$:
\be
\ba{l}
\dsp{ \frac{d^2 m^{\mu}}{d\tau^2} + 2u^{\lb} \brGam_{\lb\nu}^{\;\;\;\mu} \frac{dm^{\nu}}{d\tau} 
 + u^{\kg} u^{\lb} \der_{\nu} \brGam_{\kg\lb}^{\;\;\;\mu}\, m^{\nu} = S^{\mu}[n], }\\
 \\
\dsp{ S^{\mu}[n] \equiv - 2 \brGam_{\lb\nu}^{\;\;\;\mu} \frac{dn^{\lb}}{d\tau} \frac{dn^{\nu}}{d\tau} - 4 \der_{\kg} \brGam_{\lb\nu}^{\;\;\;\mu}\, 
u^{\lb} n^{\kg} \frac{dn^{\nu}}{d\tau} - \der_{\sg} \der_{\kg} \brGam_{\lb\nu}^{\;\;\;\mu} u^{\lb} u^{\nu} n^{\sg} n^{\kg}. }
\ea
\label{3.1}
\ee
As expected from eq.\ (\ref{1.4}) the left-hand side of this equation is identical to the first-order equation (\ref{2.0}). 
Thus an arbitrary solution of eq.\ (\ref{2.0}) can be added to any solution of (\ref{3.1}). The right-hand side of these 
equations is a quadratic expression in the first-order solutions and their derivatives. As in general the first-order 
solution is a combination of periodic and secular terms, the inhomogeneous terms $S^{\mu}[n]$ will contain various 
products of periodic and secular terms. For example, considering again the case of a reference 
geodesic along which the metric is constant, these inhomogeneous terms will be of the form
\be
S^{\mu}[n] = A_c^{\mu} \cos 2 \og \tau + A_s^{\mu} \sin 2 \og \tau + B_c^{\mu} \cos \og \tau + B_s^{\mu} \sin \og \tau + C^{\mu},
\label{4.0}
\ee
where $\og$ is an eigenfrequency of the linear operator acting on $m^{\mu}$ on the left-hand side of the first equation (\ref{3.1}). 
When the coefficients $B^{\mu}_{c, s}$ are non-zero we have resonant driving forces which lead to singular results for
the amplitude of $m^{\mu}$. 

The origin and resolution of this kind of singular behaviour in the perturbative treatment of non-linear oscillators
was realized long ago by Lindstedt and Poincar\'{e} \ct{lind, poinc} (for a modern presentation, see e.g.\ 
\ct{jorge-saletan}). Briefly, the dependence of the frequency on the amplitude of an anharmonic oscillator is
not properly taken into account by the naive perturbative treatment. An improved perturbation theory can be
developed in which both the amplitude and the frequencies of the perturbative solutions are made to 
depend on the expansion parameters, so as to cancel singular behaviour of the final solutions. 

In the present case the procedure is a little more involved, as we have to solve perturbative equations of 
motion (geodesics) subject to a constraint: the normalization condition $u^2 = -1$. Therefore we proceed as
follows. We start from the given reference geodesic $\bar{x}^{\mu}(\tau)$, its tangent vector $u^{\mu}(\tau)$
and the metric on the reference geodesic $\bar{g}_{\mu\nu} = g_{\mu\nu}(\bar{x})$ and its derivatives as
given. Then we develop the series expansion of neighboring geodesics $x^{\mu}[\tau;\sg]$ as in eq.\ (\ref{1.2});
both curves $x^{\mu}(\tau)$ and $\bar{x}^{\mu}(\tau)$ being geodesics, substitution of this expression  
into eq.\ (\ref{1.1}) gives
\be
\ba{ll}
0  = \dsp{ \frac{D^2 x^{\mu}}{D\tau^2} \hs{-.4} }& = \dsp{ \sg \left[ \frac{d^2 n^{\mu}}{d\tau^2} 
 + 2 u^{\lb} \brGam_{\lb\nu}^{\;\;\;\mu} \frac{dn^{\nu}}{d\tau} + 
 u^{\lb} u^{\kg} \der_{\nu} \brGam_{\kg\lb}^{\;\;\;\mu} n^{\nu} \right]_{\bar{x}} }\\
 & \\
 & \dsp{ \hs{.7} +\, \frac{1}{2} \sg^2 \left[  \frac{d^2 m^{\mu}}{d\tau^2} + 2 u^{\lb} \brGam_{\lb\nu}^{\;\;\;\mu} \frac{dm^{\nu}}{d\tau} +
 u^{\lb} u^{\kg} \der_{\nu} \brGam_{\kg\lb}^{\;\;\;\mu} m^{\nu}  - S^{\mu}[n] \right]_{\bar{x}} + ... }\\
\ea
\label{4p.1}
\ee
Now define the new time variable $\lb$ by
\be
\og \lb = \bar{\og} \tau \equiv \lh \og + \sg \og_1 +  \sg^2 \og_2 + ... \rh \tau,
\label{4p.2}
\ee
where $\og$ is the eigenfrequency characterizing the first-order deviation, and $\og_{1, 2, ...}$ are 
higher-order corrections. The expansion then takes the equivalent form
\be
\ba{lll}
 0 & = & \dsp{ \sg \left[ \frac{d^2 n^{\mu}}{d\lb^2} + 2 u^{\lb} \brGam_{\lb\nu}^{\;\;\;\mu} 
 \frac{dn^{\nu}}{d\lb} + u^{\kg} u^{\lb} \der_{\nu} \brGam_{\kg\lb}^{\;\;\;\mu} n^{\nu} \right]_{\bar{x}} }\\
 & & \\
 & & \dsp{ +\, \frac{1}{2}\, \sg^2 \left[ \frac{d^2 m^{\mu}}{d\lb^2} + 2 u^{\lb} \brGam_{\lb\nu}^{\;\;\;\mu} 
 \frac{dm^{\nu}}{d\lb} + u^{\kg} u^{\lb} \der_{\nu} \brGam_{\kg\lb}^{\;\;\;\mu} m^{\nu} - \Sg^{\mu}[n] \right]_{\bar{x}} + ..., }
\ea
\label{4p.3}
\ee
where the inhomogeneous source term for $m^{\mu}$ is changed to
\be
\ba{lll}
\Sg^{\mu}[n] & = & \dsp{ - 2 \brGam_{\lb\nu}^{\;\;\;\mu} \frac{dn^{\lb}}{d\lb} \frac{dn^{\nu}}{d\lb} 
 - 4 \der_{\kg} \brGam_{\lb\nu}^{\;\;\;\mu}\, u^{\lb} n^{\kg} \frac{dn^{\nu}}{d\lb} 
 - \der_{\sg} \der_{\kg} \brGam_{\lb\nu}^{\;\;\;\mu} u^{\lb} u^{\nu} n^{\sg} n^{\kg} }\\
 & & \\
 & & \dsp{ -\,\frac{4 \og_1}{\og}\, \frac{d^2 n^{\mu}}{d\lb^2} - \frac{4 \og_1}{\og}\, \brGam_{\lb\nu}^{\;\;\;\mu} u^{\lb} \frac{dn^{\nu}}{d\lb}. }
\ea
\label{4p.4}
\ee
It is now possible to choose $\og_1$ and its higher-order generalizations so as to cancel the dangerous 
contributions in the inhomogeneous terms  $\Sg^{\mu}[n]$ which produce the divergences. Observe, that 
this is achieved by a rearrangement of the perturbative expansion of the geodesics $x^{\mu}[\tau;\sg]$.
In the next section we illustrate the procedure for the example of bound geodesics in Schwarzschild space-time.

\section{Second-order geodesics in Schwarzschild space-time \label{s5}}

In this section we construct the solutions to the second-order deviation equations for bound orbits in Schwarzschild 
space-time. However, we must first briefly return to the first-order deviations. The equation for the first-order deviations
is the linear homogeneous equation
\be
\frac{d^2 n^{\mu}}{d\lb^2} + 2 u^{\lb} \brGam_{\lb\nu}^{\;\;\;\mu} 
 \frac{dn^{\nu}}{d\lb} + u^{\kg} u^{\lb} \der_{\nu} \brGam_{\kg\lb}^{\;\;\;\mu} n^{\nu} = 0,
\label{5.1}
\ee
with regular periodic solutions 
\be
n_{per}^{\mu}(\lb) = n_c^{\mu} \cos \og \lb + n_s^{\mu} \sin \og \lb.
\label{5.2}
\ee
Substitution of this expression in eq.\ (\ref{5.1}) returns eqs.\ (\ref{2.7}), showing that the first-order geodesic deviations
of a circular orbit are unchanged in the rearranged perturbation theory, except for changing $\tau \rightarrow \lb$:
\be
\ba{lll}
n^t(\lb) & = & \dsp{ - \frac{\ag}{\og}\, n_c^r \sin \og \lb + \frac{\ag \kg \lb}{\ag \bg - \gam \eta}\, \Del_n, }\\
 \\
n^r(\lb) & = & \dsp{ n^r_c \cos \og \lb + \Del_n, }\\
 \\
n^\vf(\lb) & = & \dsp{ - \frac{\eta}{\og}\, n_c^r \sin \og \lb  + \frac{\eta \kg \lb}{\ag \bg - \gam \eta}\, \Del_n,}
\ea
\label{5.3}
\ee
with $\og \lb = \bar{\og} \tau$, showing that the proper frequency has been changed by terms of order $\sg$ and higher. 
Also note the relation
\be
\brg_{\mu\nu} u^{\mu} n^{\nu} = - \ve_0 n^t + \ell_0 n^{\vf} = 0.
\label{5.3.1}
\ee
It follows directly, that  $n^{\mu}(\lb)$ is an exact solution of the condition (\ref{2.13})
\be
\brg_{\mu\nu} u^{\mu} \frac{Dn^{\nu}}{D\tau} = \brg_{\mu\nu} u^{\mu} \lh \frac{d\lb}{d\tau} \frac{dn^{\nu}}{d\lb} 
 + u^{\lb} \brGam_{\lb\kg}^{\;\;\;\nu} n^{\kg} \rh =  \brg_{\mu\nu} u^{\mu}\, \frac{Dn^{\nu}}{D\lb} \frac{d\lb}{d\tau} = 0,
\label{5.3.2}
\ee
where the pull back of the covariant derivative is
\be
\frac{Dn^{\mu}}{D\lb} = \frac{dn^{\mu}}{d\lb} + \frac{d\bar{x}^{\lb}}{d\lb} \brGam_{\lb\nu}^{\;\;\;\mu} n^{\nu} 
 = \frac{dn^{\mu}}{d\lb} + \frac{d\tau}{d\lb} u^{\lb} \brGam_{\lb\nu}^{\;\;\;\mu} n^{\nu}.
\label{5.3.3} 
\ee
Substitution of the solutions (\ref{5.3}) in the inhomogeneous terms (\ref{4p.4}) of the second-order deviation equations 
now provides expressions for $\Sg^{\mu}$ of the form
\be
\ba{lll}
\Sg^t & = & \dsp{ a^t \sin 2 \og \lb + b^t \sin \og \lb, }\\
 & & \\
\Sg^r & = & \dsp{ a^r \cos 2\og \lb + b^r \cos \og \lb + c^r, }\\
 & & \\
\Sg^{\vf} & = & \dsp{ a^{\vf} \sin 2 \og \lb + b^{\vf} \sin \og \lb. }
\ea
\label{5.4}
\ee
The coefficients $(a^{\mu}, b^{\mu}, c^{\mu})$ are complicated expressions in terms of $R$ and the
first-order deviation parameters $(n_c^r, \Del_n)$, which are given in appendix \ref{sab}. Then the solutions 
for $m^{\mu}$ take much the same form, with additional secular terms
\be
\ba{lll}
m^t & = & \dsp{ m_2^t \sin 2 \og \lb + m_1^t \sin \og \lb + w^t \lb, }\\
 & & \\
m^r & = & \dsp{ m_2^r \cos 2\og \lb + m_1^r \cos \og \lb + \Del_m, }\\
 & & \\
m^{\vf} & = & \dsp{ m_2^{\vf} \sin 2 \og \lb + m_1^{\vf} \sin \og \lb + w^{\vf} \lb, }
\ea
\label{5.5}
\ee
where the coefficients are solutions of the linear systems 
\be
\lh \ba{ccc} - 4 \og^2 & - 2 \og \ag & 0 \\
                 2 \og \bg & - (4 \og^2 + \kg) & - 2 \og \gam \\
                 0 & - 2 \og \eta & - 4 \og^2 \ea \rh 
                 \lh \ba{c} m_2^t \\ m_2^r \\ m_2^{\vf} \ea \rh 
              = \lh \ba{c} a^t \\ a^r \\ a^{\vf} \ea \rh,           
\label{5.6}
\ee
\be
\lh \ba{ccc} - \og^2 & - \og \ag & 0 \\
                 \og \bg & - (\og^2 + \kg) & - \og \gam \\
                 0 & - \og \eta & - \og^2 \ea \rh 
                 \lh \ba{c} m_1^t \\ m_1^r \\ m_1^{\vf} \ea \rh 
              = \lh \ba{c} b^t \\ b^r \\ b^{\vf} \ea \rh,  
\label{5.7}
\ee
and
\be
\bg w^t - \gam w^{\vf} - \kg \Del_m = c^r.
\label{5.8}
\ee
The subtlety in solving these equations is, that the determinant of the matrix of coefficients of the $m_1^{\mu}$ vanishes
as a result of the relation (\ref{2.8}). The system of equations (\ref{5.7}) can be solved only, if the inhomogeneous terms
$b^{\mu}$ have vanishing components in the direction of the zero mode of the coefficient matrix; for this to happen, the 
frequency shift $\og_1$ has to be chosen properly. 

To be precise, for $\og \neq 0$ the coefficient matrix on the left-hand side of eq.\ (\ref{5.7}) has three different eigenvalues:
$(0, - \og^2, - (2\og^2 + \kg))$. The  left zero mode is given (up to an irrelevant normalization factor) by
\be
m_{0\, \mu} = \lh \bg, \og, -\gam \rh.
\label{5.9}
\ee
Therefore the condition that eq.\ (\ref{5.7}) is invertible is
\be
\bg b^t + \og b^r - \gam b^{\vf} = 0. 
\label{5.10}
\ee
Now all coefficients $b^{\mu}$ are of the form
\be
b^{\mu} = n_c^r \lh F^{\mu}\, \frac{\Del_n}{R} - G^{\mu}\, \frac{\og_1}{\og} \rh,
\label{5.11}
\ee
with $(F^{\mu}, G^{\mu})$ as given in appendix \ref{sab}, determined only by $M$ and $R$. As a result we finally obtain
\be
\frac{\og_1}{\og} = \frac{\bg F^t + \og F^r - \gam F^{\vf}}{ \bg G^t + \og G^r - \gam G^{\vf}}\, \frac{\Del_n}{R}.
\label{5.12}
\ee
It should be noted, that the frequency shift $\og_1$ is proportional to the secular radial shift $\Del_n$, but independent
of $n_c^r$, and hence vanishes whenever $\Del_n = 0$. Provided the constraint (\ref{5.10}) on the inhomogeneous terms
is satisfied, eq.\ (\ref{5.7}) can be inverted to yield
\be
\ba{l}
\dsp{ m_1^t = - \frac{\eta b^t - \ag b^{\vf}}{\og^2(\og^2 + \kg)}\, \gam + \frac{\bg b^t - \gam b^{\vf}}{(2\og^2 + \kg)(\og^2 + \kg)}\, \ag, }\\
 \\
\dsp{ m_1^r = \frac{\bg b^t - \gam b^{\vf}}{\og (2\og^2 + \kg)}, }\\
 \\
\dsp{ m_1^{\vf} = - \frac{\eta b^t - \ag b^{\vf}}{\og^2(\og^2 + \kg)}\, \bg + \frac{\bg b^t - \gam b^{\vf}}{(2\og^2 + \kg)(\og^2 + \kg)}\, \eta.}
\ea
\label{5.13}
\ee
By construction these coefficients $m_1^{\mu}$ satisfy the same constraint as the source terms $b^{\mu}$:
\be
\bg m_1^t + \og m_1^r - \gam m_1^{\vf} = 0,
\label{5.13.1}
\ee
i.e.\ they are also orthogonal to the zero-mode (\ref{5.9}). 
In contrast, eqs.\ (\ref{5.6}) for $m_2^{\mu}$ can be inverted straightforwardly, with the result
\be
\ba{l}
\dsp{ m_2^t = \frac{\ag}{12\og^4} \lh \bg a^t + 2 \og a^r - \gam a^{\vf} \rh - \frac{a^t}{4 \og^2}, }\\
 \\
\dsp{ m_2^r = - \frac{1}{6\og^3} \lh \bg a^t + 2 \og a^r - \gam a^{\vf} \rh, }\\
 \\
\dsp{ m_2^{\vf} = \frac{\eta}{12 \og^4} \lh \bg a^t + 2 \og a^r - \gam a^{\vf} \rh - \frac{a^{\vf}}{4 \og^2}. }
\ea
\label{5.14}
\ee
Of course, these solutions are determined only up to a solution $\tilde{m}^{\mu}$ of the homogeneous equation, depending
on the initial conditions. 

As concerns the normalization of the four-velocity, in stead of eq.\ (\ref{2.13}) the solutions up to second order must now 
satisfy the condition
\be
\sg \lh u_{\mu} \frac{Dn^{\mu}}{D\tau} \rh + \frac{\sg^2}{2} \lh u_{\mu} \frac{Dk^{\mu}}{D\tau} 
 + \bar{g}_{\mu\nu} \frac{Dn^{\mu}}{D\tau} \frac{Dn^{\nu}}{D\tau} + u^{\kg}u^{\lb} \bar{R}_{\kg\mu\lb\nu} n^{\mu} n^{\lb} \rh = 0,
\label{5.15}
\ee
where $k^{\mu} = m^{\mu} + \brGam_{\lb\nu}^{\;\;\;\mu} n^{\lb} n^{\nu}$. We have already seen in eq.\ (\ref{5.3.2}) that the first term
vanishes identically if we substitute the solution (\ref{5.3}) for $n^{\mu}(\lb)$.  Upon using the first equation (\ref{1.4})
the term of second order in $\sg$ in (\ref{5.15}) vanishes if
\be
u_{\mu} \frac{Dk^{\mu}}{D\tau} + \frac{1}{2} \frac{d^2 n^2}{d\tau^2} = 
 \frac{d}{d\tau} \lh u_{\mu} k^{\mu} + \frac{1}{2} \frac{dn^2}{d\tau} \rh = 0,
\label{5.16}
\ee
where $n^2 = \bar{g}_{\mu\nu} n^{\mu} n^{\nu}$. But to this fixed order in $\sg$ we are free to replace the proper time by the 
evolution parameter $\lb$. After substitution of the explicit expressions for $u^{\mu}$, $n^{\mu}$ and $k^{\mu}$ eq.\ (\ref{5.16}) 
then reduces to
\be
\ba{lll}
\dsp{ \ve_0\, \frac{dm^t}{d\lb} - \ell_0\, \frac{dm^{\vf}}{d\lb} }& = & \dsp{ 2 \og \lh \ve_0 m_2^t - \ell_0 m_2^{\vf} \rh \cos 2 \og \lb 
 + \og \lh \ve_0 m_1^t - \ell_0 m_1^{\vf} \rh \cos \og \lb + \ve_0 w^t - \ell_0 w^{\vf} }\\
 & & \\
 & = &\dsp{ -\, \frac{\og^2(n^r_c)^2}{1 - \frac{2M}{R}}\,  \cos 2 \og \lb
 - \frac{2 \og^2  \Del_n n^r_c}{1 - \frac{2M}{R}}\, \cos \og \lb - \frac{3}{4}\, \frac{\og^2(\Del_n)^2}{1 - \frac{3M}{R}}. }
\ea
\label{5.18}
\ee 
The two relations that follow for $m_{1,2}^{\mu}$ by comparing the terms proportional to $\cos \og \lb$ and $\cos  2 \og \lb$ 
are identities, implied by eqs.\ (\ref{5.13}) and (\ref{5.14}); however, for the constant terms we find a constraint
\be
\ve_0 w^t - \ell_0 w^{\vf} = - \frac{3}{4}\,\frac{\og^2 (\Del_n)^2}{1 - \frac{3M}{R}}.
\label{5.19}
\ee
Together with the relation (\ref{5.8}) this can be used to express $w^t$ and $w^{\vf}$ in terms of $\Del_m$ and the
lower-order parameters:
\be
\ba{lll}
w^t & = & \dsp{ - \frac{3}{4} \frac{\og^2 (\Del_n)^2}{\lh 1 - \frac{3M}{R} \rh^{3/2}} - \frac{1}{2} \frac{ R \lh c^r + \kg \Del_m \rh}{
 \lh 1 - \frac{2M}{R} \rh \sqrt{1 - \frac{3M}{R}}}, }\\
 & & \\
w^{\vf} & = & \dsp{ - \frac{3}{4R} \sqrt{\frac{M}{R}}  \frac{\og^2 (\Del_n)^2}{\lh 1 - \frac{3M}{R} \rh^{3/2}} 
 - \frac{1}{2} \sqrt{\frac{R}{M}}\, \frac{(c^r + \kg \Del_m)}{\sqrt{1 - \frac{3M}{R}}}, }
\ea
\label{5.20}
\ee
with $c^r$ as given in eq.\ (\ref{ab.11}), and $\Del_m$ a new free parameter to be fixed by the boundary conditions on $r$.
More specifically, the net result for the geodesic solutions to second order is
\be
\ba{lll}
t & = & U_0^t \tau + U_1^t \sin \bar{\og} \tau + U_2^t \sin 2 \bar{\og}\tau, \\
 & & \\
r & = & U_0^r + U_1^r \cos \bar{\og} \tau + U_2^r \cos 2 \bar{\og} \tau, \\
 & & \\
\vf & = & U_0^{\vf} \tau + U_1^{\vf} \sin \bar{\og} \tau + U_2^{\vf} \sin 2 \bar{\og} \tau, 
\ea
\label{5.21}
\label{G61}
\ee
where to this order $\bar{\og} = \og + \sg \og_1$, and
\be
\ba{l}
\dsp{ U_0^t = \frac{1}{\sqrt{1 - \frac{3M}{R}}} + \frac{\sg \ag \kg \bar{\og} \Del_n}{\og (\ag \bg - \gam \eta)} + \frac{\sg^2 \bar{\og} w^t}{2\og}, 
\hs{2} U_1^t = - \frac{\sg \ag n^r_c}{\og} + \frac{1}{2}\, \sg^2 m_1^t, \hs{2} U_2^t = \frac{1}{2}\, \sg^2 m_2^t, }\\
 \\
\dsp{ U_0^r = R + \sg \Del_n + \frac{1}{2}\, \sg^2 \Del_m, \hs{2}
 U_1^r = \sg n^r_c + \frac{1}{2}\, \sg^2 m_1^r, \hs{2} U_2^r = \frac{1}{2}\, \sg^2 m_2^r, }\\
 \\ 
\dsp{ U_0^{\vf} = \frac{1}{R} \sqrt{\frac{M}{R}} \frac{1}{\sqrt{ 1 - \frac{3M}{R}}} + \frac{\sg \eta \kg \bar{\og} \Del_n}{\og (\ag \bg - \gam \eta)}
 + \frac{\sg^2 \bar{\og} w^{\vf}}{2 \og}, \hs{.5}
 U_1^{\vf} = - \frac{\sg \eta n^r_c}{\og} + \frac{1}{2}\, \sg^2 m_1^{\vf}, \hs{.5} U_2^{\vf} = \frac{1}{2}\, \sg^2 m_2^{\vf}. }
\ea
\label{5.22}
\label{G62}
\ee
The second equation (\ref{5.21}) implies, that again the periastra and apastra occur at proper times $\bar{\og} \tau_n = n \pi$. 
It follows that
\be
r_{pa} = U_0^r + U_1^r + U_2^r, \hs{2} r_{aa} = U_0^r - U_1^r + U_2^r.
\label{5.23}
\ee
Then the equations (\ref{2.15.2}) are modified to
\be
\ba{l}
\dsp{ \frac{1}{2} \lh r_{pa} - r_{aa} \rh = \sg n_c^r + \frac{1}{2}\, \sg^2 m_1^r, }\\
 \\
\dsp{ \frac{1}{2} \lh r_{pa} + r_{aa} \rh - R = \sg \Del_n + \frac{1}{2}\, \sg^2 \lh \Del_m + m_2^r \rh, }
\ea
\label{5.24}
\ee
where still relation (\ref{2.9.1}) holds between $n_c^r$ and $\Del_n$. Hence these equations determine $\Del_n$ and $\Del_m$
in terms of the observables $(r_{pa}, r_{aa})$, and the expansion parameter $\sg$:
\be
\sg \Del_n - \frac{1}{2} \sg^2 m_1^r = \frac{1}{2} \lh r_{aa} - r_{pa} \rh + \sg \sqrt{1 - \frac{2M}{R}}, \hs{2} 
\frac{1}{2} \sg^2 \lh \Del_m + m_1^r + m_2^r \rh = r_{pa} - R- \sg \sqrt{1 - \frac{2M}{R}}.
\label{5.24.1}
\ee
The parameter $\sg$ itself is fixed by the time or angle between periastra, which can be expressed in terms of the 
proper-time lapse between periastra: 
\be
\bar{\og} \tau_2 = \lh \og + \sg \og_1 \rh \Del \tau = 2 \pi.
\label{5.25}
\ee
As in the first-order approximation we can fix the constants of motion $\ve$ and $\ell$ by evaluating the expressions
(\ref{aa.2}) after substitution of (\ref{5.21}) to second order in $\sg$; the result is 
\be
\ba{lll}
\ve_m & = & \dsp{ \left[ \lh 1 - \frac{2M}{U_0^r} \rh U_0^t - \frac{MU_0^t \left( U_1^r \right)^2}{\left( U_0^r \right)^3} 
 + \frac{M \bar{\og} U_1^t U_1^r}{\left( U_0^r \right)^2} \right]_{\sg^2}, }\\
 & & \\
\ell_m & = & \left[ U_0^{\vf} \left( U_0^r \right)^2 + \frac{1}{2}\, U_0^{\vf} \left( U_1^r \right)^2 
 + \bar{\og}\, U_0^r U_1^r U_1^{\vf} \right]_{\sg^2},
\ea
\label{5.26}
\label{G67}
\ee
where the notation implies that the expressions are to be truncated at order $\sg^2$.

\section{Numerical results}

In the previous sections we have constructed a covariant perturbation theory for orbits of test masses in curved space-time
based on the method of geodesic deviations, which provides an alternative to the common post-newtonian expansions. 
We have applied it in particular to motion in a Schwarzschild background geometry, and obtained explicit expressions for 
bound orbits to second order in the geodesic deviation parameter. 
Using these expressions for the orbits, we can now investigate how well the resulting epicycle orbits compare with the 
ones that are obtained when the geodesic equations are solved purely numerically. As advocated, the main advantage of the 
geodesic deviation method is the fact that the curvature of the spacetime is taken into full account, and hence it is expected that 
the orbits will remain accurate even when considered very close to the black hole. In order to test this, two explicit examples will 
be considered of orbits in the region of extremely curved spacetime, \emph{i.e.} that have their periasta close to $6M$, the radius 
of the innermost stable circular orbit (ISCO). The comparison to the purely numerically calculated orbits will be done 
for both the epicycle expressions up to first order, and up to second order. 
\newline
\newline
For the first example the mass of the black hole is set to $M=10$ in some unspecified units. In the same units, we take the 
following periastron distance $r_{pa}$, apastron distance $r_{aa}$, periastron shift $\delta \varphi$ and proper timeshift $\Delta \tau$ 
between successive periastra:
\be
r_{pa} = 63.1228, \quad \quad \quad r_{ap} = 92.5279,\quad \quad \quad \delta \varphi = 7.92000, \quad \quad \quad \Delta \tau = 2984.06
\ee
(It should be noted that the fact that this orbit has a periastron shift greater than the Newtonian angular distance between periastra 
of $2\pi$, is evidence of the very strong relativistic effect this close to the central mass.) This orbit is uniquely fixed
by the following values of the energy and angular momentum per unit mass of the test particle:
\be
\varepsilon = 0.94827 \quad \quad \quad \quad \ell = 35.5000.
\label{G69}
\ee
This orbit will now be constructed up to first and up to second order in epicycle perturbation theory. In the case of the
 first-order orbits, eq.\ (\ref{G26}), there are three constants that need to be assigned values: $\sg, R, \Delta_n$. 
This means that these functions can be subjected to three boundary conditions to fix the orbit. The following will be used: 
the orbit must have a periastron shift $\delta \varphi$, have two successive periastra a proper time $\Delta \tau$ apart, 
and yield a radial periastron distance $r_{pa}$. From eq.\ (\ref{G26}), these three conditions mathematically translate to 
the algebraic conditions
\begin{eqnarray}
\delta \varphi &=& 2 \pi \left( \frac{1}{\omega} \sqrt{\frac{M}{R^3}} \frac{1}{\sqrt{1-\frac{3M}{R}}} + 
\frac{1}{\omega}\frac{\eta \kappa}{\alpha \beta - \gamma \eta}\sigma \Delta_n  -1 \right) \nonumber \\
\Delta \tau &=& \frac{2 \pi}{\omega} \nonumber \\
r_{pa} &=& R+\sigma \lh n_c^r +  \Delta_n \rh 
\end{eqnarray}
Solving the three boundary conditions then yields the following values for the normalisation parameter $\sigma$ and the 
independent epicycle parameters $R, \Delta_n$ as well as the associated value of $n_c^r$:
\be
\sigma = -25.9415, \quad \quad R = 85.8422, \quad \quad \Delta_n = 0.449635, \quad \quad n_c^r = 0.426159,
\ee
and hence the following expressions for the orbital functions $t(\tau), r(\tau), \varphi(\tau)$:
\begin{eqnarray}
t(\tau) &=& 1.2851 \, \tau + 17.9319 \, \sin(\omega \tau) \nonumber \\
r(\tau) &=& 74.1780 -11.0552 \,\cos(\omega \tau) \nonumber \\
\varphi (\tau) &=& 0.00611 \, \tau + 0.46944 \, \sin(\omega \tau)
\end{eqnarray}
in which the epicycle frequency is given by 
\be
\omega = 0.00270.
\ee
Using the expressions for the $\varepsilon_n$ and $\ell_n$ from eqs.\ (\ref{2.17}) and (\ref{G30}), the energy per unit 
mass and the angular momentum per unit mass of this epicycle orbit are given by
\be
\varepsilon_n = 0.94644, \quad \quad \quad \ell_n = 35.1840,
\ee
which are at most a few tenths of a per cent different from the values in eq.\ (\ref{G69}). 

Moving on to the second order, the orbital functions eq.\ (\ref{G62}) can be subjected to the same three boundary conditions as before, 
but this time an extra constant $\Delta_m$ appears that needs to be assigned a value. This means that a fourth boundary 
condition can be imposed, which will be chosen to fit the radial apastron distance $r_{aa}$. The four boundary conditions 
are formulated as algebraic conditions by observing that the periastra and apastra 
correspond to the extreme values of $r(\tau)$ in eq.\ (\ref{G61}) and that these occur at times $\bar{\omega} \tau_n = n \pi$, 
$n$ an integer number. From $\varphi(\tau)$ then follows the periastron shift, while $r(\tau)$ yields the radial distances of 
the periastra and apastra. The four boundary conditions then translate to the following algebraic conditions:
\begin{eqnarray}
\Delta \tau = \frac{2 \pi}{\omega+\sigma \omega_1} &,& \quad \quad \delta \varphi = 
 2 \pi \left(\frac{U_0^\varphi }{\omega+\sigma \omega_1}-1 \right), \nonumber \\
r_{pa} = U_0^r+U_1^r+U_2^r &,& \quad \quad r_{aa} = U_0^r-U_1^r+U_2^r. 
\end{eqnarray}
Solving these conditions then yields values for the normalisation parameter $\sigma$ and the independent epicycle 
parameters $R, \Delta_n$, $\Delta_m$ as well as the associated value of $n_c^r$. However, 
when comparing the epicycle approximation in higher order to a specific known orbit, there is generically more than one 
set of values $\sigma, R, \Delta_n, \Delta_m$ solving the boundary conditions. The question then presents itself which of 
these orbits is the most accurate one to describe the (unique) solution to the geodesic equation parametrized by the values 
$\varepsilon, \ell$. The answer is provided by the previously made observation that the expansion parameter of the 
epicycle approximation is the dimensionless number $\sigma n_c^r /R$, and hence that the solution set that produces 
the smallest value of this parameter yields the most accurate orbit. Using this criterium, the following solution set 
$\sigma, R, \Delta_n, \Delta_m$ and associated value $n_c^r$ is chosen:
\be
\sigma = -3.11032, \quad  R = 79.8763,  \quad \Delta_n = -2.88620,  \quad \Delta_m = -2.19232,  \quad n_c^r = 3.75194,
\ee
which give rise to the following expressions for the orbital functions
\begin{eqnarray}
t(\tau) &=& 1.28429 \, \tau + 22.0076 \, \sin((\omega + \sigma \omega_1)\tau) + 1.47690 \, \sin(2(\omega + \sigma \omega_1)\tau),  \nonumber \\
r(\tau) &=& 78.2490 \, -14.7025 \,\cos((\omega+\sigma \omega_1) \tau) -0.42359 \,\cos(2(\omega+\sigma \omega_1) \tau),   \nonumber \\
\varphi (\tau) &=& 0.00611 \, \tau + 0.57279 \, \sin((\omega+\sigma \omega_1) \tau) + 0.04272 \, \sin(2(\omega+\sigma \omega_1) \tau), \nonumber
\end{eqnarray}
in which the frequencies are given by
\be
\omega = 0.00279, \quad \quad \sigma \omega_1 = -0.00009, \quad \quad \omega+\sigma \omega_1 = 0.00270.
\ee
From eq.\ (\ref{G67}), the energy per unit mass and angular momentum per unit mass for this approximation to the orbit then follow as
\be
\varepsilon_m = 0.94807, \quad \quad \quad \ell_m  = 35.5802,
\ee
which are at most a few hundredths of a per cent different from the values in eq.(\ref{G69}). \newline
\newline
Having found the epicycle expressions to first and second order, they can be compared to orbital functions $(t(\tau), r(\tau), \vf(\tau))$ 
as calculated by solving the geodesic equations by purely numerical means. The latter are completely specified by the values 
$\varepsilon$ and $\ell$ of eq.\ (\ref{G69}) by methods explained in \ct{Cutler}.

\begin{figure}[!h]
  \begin{center}
    \includegraphics[scale=0.8]{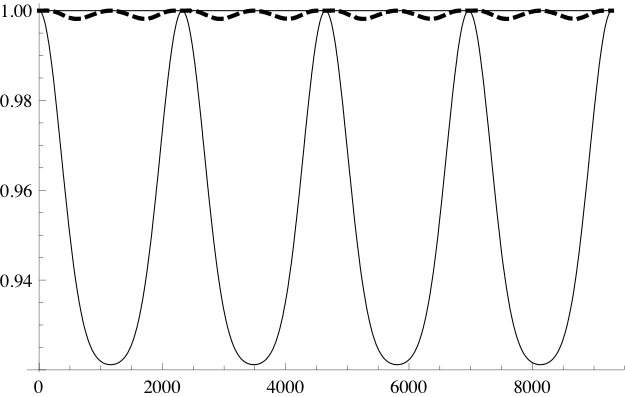}
  \end{center}
    \caption{The radial function $r(\tau)$ in epicycle approximation up to first (solid line) and second (dashed line) order divided by the 
    numerical one, as a function of proper time $\tau$. This is the result for the first example mentioned in the text. }
    \label{FigureR75}
\end{figure}

\begin{figure}[!h]
  \begin{center}
    \includegraphics[scale=0.8]{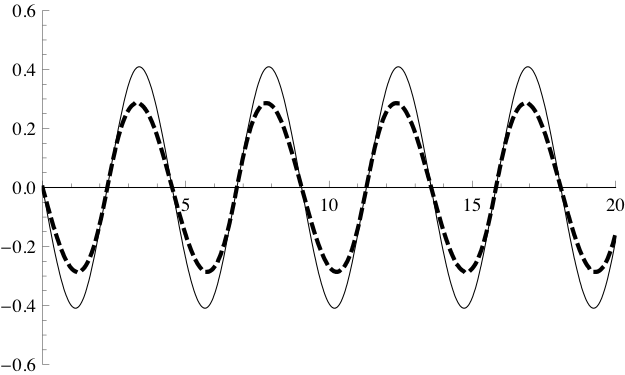}
  \end{center}
    \caption{The difference of the angular function $\varphi(\tau)$ in epicycle approximation up to first (solid line) and second (dashed line) order minus 
    the numerical one, as a function of the numerical $\varphi$ in units of $\pi$.This is the result for the first example mentioned in the text. }
    \label{FigurePhi75}
\end{figure}

In Figure \ref{FigureR75}, the radial function $r(\tau)$ up to first and second order is given divided by its purely numerical counterpart, 
as a function of proper time. As can be seen, the relative difference between our first-order approximation and the numerical one is 
at most about $8\%$; introducing the second-order epicycle improves the relative difference to less than $0.5 \%$. Figure \ref{FigurePhi75}
shows the absolute difference between the angular co-ordinate $\varphi$ in the epicycle approximation (both to first and 
second order) and in the numerical one. As can be seen, the first-order epicycle deviates from the purely numerical one by at 
most $0.4$ radians during any period, whereas the second order deviates from the numerical one by at most $0.25$ radians.

The figures show that the radial coordinate $r(\tau)$ is very well approximated by the second-order epicycle functions. On the other 
hand, in the second-order epicycle approximation the coordinates $\varphi(\tau)$ and $t(\tau)$ still show some small deviations; 
further improvement could be made by including third- and higher-order corrections.

As a second and less extreme example, the mass of the black hole will again be set to $M = 10$, and an orbit will be considered 
which has, in the same units, the following periastron distance $r_{pa}$, apastron distance $r_{aa}$, periastron 
shift $\delta \varphi$ and proper timeshift $\Delta \tau$ between successive periastra:
\be
r_{ap} = 100.000, \quad \quad r_{aa} = 150.000, \quad \quad \delta \varphi = 2.61000, \quad \quad \Delta \tau = 3373.56.
\ee
This orbit is bound between $10 M$ and $15 M$, and corresponds to one that has energy per unit mass and angular momentum 
per unit mass given by:
\begin{equation}
\varepsilon = 0.96362, \quad \quad \quad \quad \ell = 40.0892.
\label{G80}
\end{equation}
Subjecting the first order epicycles of eq.\ (\ref{G26}) to boundary conditions similar to the ones discussed in the previous example, 
the values for the normalisation parameter $\sigma$ and the epicycle parameters $R,  \Delta_n$ as well as the associated value 
$n_c^r$ follow as:
\be
R = 125.515, \quad \quad \sigma = -27.8282,  \quad \quad \Delta_n = 0.06148, \quad \quad n_c^r = 0.85540,
\ee
which give the following first order orbital functions:
\begin{eqnarray}
t(\tau) &=& 1.14879 \, \tau + 22.1255 \, \sin(\omega \tau), \nonumber \\
r(\tau) &=& 123.804 - 23.8041 \,\cos(\omega \tau), \nonumber \\
\varphi (\tau) &=& 0.00264 \, \tau + 0.52501 \, \sin(\omega \tau),
\end{eqnarray}
in which the epicycle frequency is given by 
\be
\omega = 0.00186.
\ee
Using the expressions for the $\varepsilon_n$ and $\ell_n$ from eq.\ (\ref{G30}), the energy per unit mass and angular momentum 
per unit mass of this epicycle orbit are given by
\be
\varepsilon_n = 0.96325, \quad \quad \quad \ell_n = 40.4225, 
\ee
which deviate from the numerical counterparts of eq.\ (\ref{G80}) by less than a per cent. 

Going to second order eq.\ (\ref{G62}), an analysis similar to the one discussed in the previous example yields the following  values for the 
normalisation parameter $\sigma$ and the epicycle parameters $R, \, \Delta_n, \, \Delta_m$ as well as the associated value for $n_c^r$ :
\be
\sigma = -20.6911, \quad  R = 120.507,  \quad \Delta_n = -0.23446,  \quad \Delta_m = 0.00745,  \quad n_c^r = 1.14772, 
\ee
which give the following second order orbital functions:
\begin{eqnarray}
t(\tau) &=& 1.14876 \, \tau + 23.2580 \, \sin((\omega + \sigma \omega_1)\tau) + 2.31954 \, \sin(2(\omega + \sigma \omega_1)\tau),  \nonumber \\
r(\tau) &=& 126.953 \, -25.0000 \,\cos((\omega+\sigma \omega_1) \tau) -1.95325 \,\cos(2(\omega+\sigma \omega_1) \tau),   \nonumber \\
\varphi (\tau) &=& 0.00264 \, \tau + 0.55205 \, \sin((\omega+\sigma \omega_1) \tau) + 0.06398 \, \sin(2(\omega+\sigma \omega_1) \tau), \nonumber \\ 
\end{eqnarray}
in which the frequencies are given by
\be
\omega = 0.00195, \quad \quad \sigma \omega_1 = -0.00009, \quad \quad \omega+\sigma \omega_1 = 0.00186,
\ee
The energy per unit mass and the angular momentum per unit mass follow from eq.\ (\ref{G67}) as
\begin{equation}
\varepsilon_m = 0.96362, \quad \quad \quad \ell_m = 40.0822,
\end{equation}
which deviate less than a hundredth of a per cent from their purely numerical counterparts of eq.\ (\ref{G80}). 

Having found the epicycle expressions to first and second order, they can again be compared to orbital functions as calculated by 
solving the geodesic equations by purely numerical means. The latter are completely specified by the values $\varepsilon$ and 
$\ell$ of eq.\ (\ref{G80}) by methods explained in \ct{Cutler}, the result being numerical functions $r(\tau)$ and $\varphi(\tau)$. 

\begin{figure}[!h]
  \begin{center}
    \includegraphics[scale=0.8]{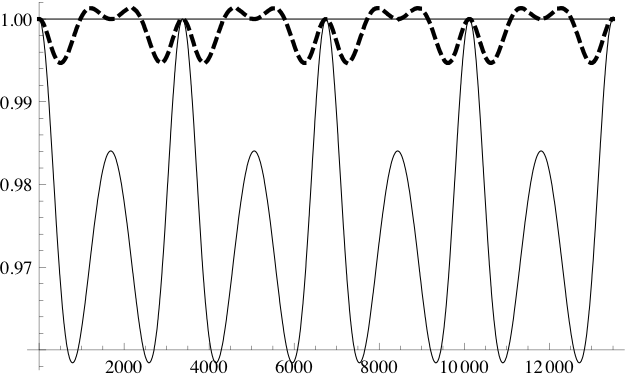}
  \end{center}
    \caption{The radial function $r(\tau)$ in epicycle approximation up to first (solid line) and second (dashed line) order divided by the 
    numerical one, as a function of proper time $\tau$. This is the result for the second example mentioned in the text. }
    \label{FigureR12}
\end{figure}

\begin{figure}[!h]
  \begin{center}
    \includegraphics[scale=0.8]{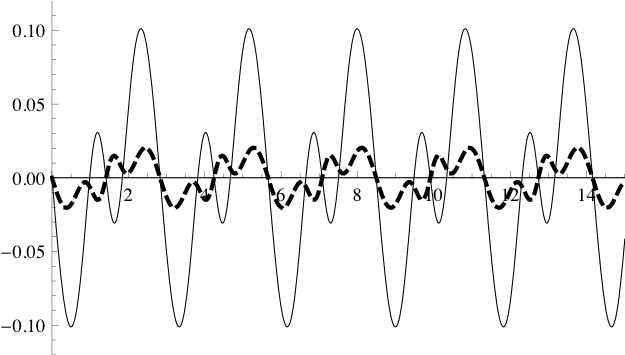}
  \end{center}
    \caption{The difference of the angular function $\varphi(\tau)$ in epicycle approximation up to first (solid line) and second (dashed line) 
    order minus the numerical one, as a function of the numerical $\varphi$ in units of $\pi$.This is the result for the second example 
    mentioned in the text. }
    \label{FigurePhi12}
\end{figure}

In Figure \ref{FigureR12}, the radial function $r(\tau)$ is shown up to first and second order as a function of proper time, compared to the purely 
numerical approximation in terms of the ratio. As can be seen, the relative difference between our first-order approximation and 
the numerical one is at most about $4\%$; introducing the second-order epicycle improves the relative difference to less than $0.5 \%$. 
Figure \ref{FigurePhi12} shows the absolute difference between the orbital function $\varphi$ (both to first and second order) and the numerical one, 
as a function of the numerical $\varphi$. As can be seen, the first-order epicycle deviates from the purely numerical one by at 
most $0.1$ radians, whilst the second-order one deviates from the numerical one by at most $0.02$ radians. These results imply that the 
orbital functions as given by the geodesic deviation method very accurately describe the geodesic orbit uniquely specified by the 
values of $\varepsilon$, $\ell$ as given in eq.(\ref{G80}). 
\vs{2}

\section{Conclusions and outlook}

The two examples presented show that the geodesic deviation method can be used to accurately describe eccentric bounded geodesics of a Schwarzschild black hole. The choice of boundary conditions used in this article focussed mostly on improving the accuracy of the radial function 
and, as such, the method effortlessly produces very accurate results for $r(\tau)$, even when describing orbits grazing the ISCO. The accuracy 
of $t(\tau)$ and $\varphi(\tau)$ can be further improved by taking the method to higher order, or for specific purposes by 
choosing boundary conditions that focus on constraining these functions rather than the radial values. 
Although each extra order will introduce an extra degree of freedom and hence allows for one extra boundary condition to be used as 
a constraint on $t(\tau)$ and $\varphi(\tau)$, it is expected that only one extra order correction suffices to improve both two functions in 
accuracy. This is because of the normalization of four-velocity and the fact that the radial orbital function is already very accurate at second 
order: the relation eq.\ (\ref{G90})
\begin{equation}
-\left( 1-\frac{2M}{r} \right) \left( \frac{dt}{d\tau} \right)^2 +\left( 1-\frac{2M}{r}\right)^{-1} \left( \frac{dr}{d\tau} \right)^2 
 + r^2 \left( \frac{d\vf}{d\tau} \right)^2 = -1
\end{equation}
will ensure an increase in accuracy of $t(\tau)$ or $\varphi(\tau)$ when the extra degree of freedom is used to improve the 
accuracy of the other.

At least two further applications of the geodesic method are immediately foreseeable. Firstly, it opens up the possibility of removing one numerical step in calculating the gravitational radiation in a Schwarzschild background. A formalism to calculate gravitational radiation due to the motion of a test mass in a Schwarzschild background has been topic of research for many decades and was presented in a final form in \ct{Martel}, \ct{Martel2}, but where that work had to rely on numerical descriptions of the geodesic orbits, the geodesic deviation method allows to replace that numerical step by an analytical one. Secondly, our presentation of the geodesic deviation method only relied on the assumption that the equations of motion of the deviation vectors were linear differential equations with constant coefficients, and the results are therefore quite general and can be used to describe other physical situations as well. For example, the method can be applied to the description of the motion of an electrically charged test mass in a Schwarzschild spacetime while experiencing a Lorentz force due to an axially symmetric magnetic field. Both of these applications will be the topic of future publications.
\newline
\newline
\nit
{\bf Acknowledgement} \\
It is a pleasure to acknowledge useful discussions with R.\ Kerner and S.\ Vitale of the Universit\'{e} Pierre et Marie Curie in Paris. 
For JWvH this work is part of the research programme of the Foundation for Fundamental Research on Matter (FOM). 

\np
\appendix

\section{Geodesics in Schwarzschild space-time \label{saa}}

In this appendix we collect some well-known result about Schwarzschild geometry which are relevant for the results derived
and discussed in the main text. We describe the static and spherically symmetric exterior geometry of Schwarzschild 
space-time in terms of the standard Schwarzschild-Droste co-ordinates (in units in which $c = G = 1$):
\be
d\tau^2 = \lh 1 - \frac{2M}{r} \rh dt^2 - \frac{dr^2}{1 - \frac{2M}{r}} - r^2 \lh d\thg^2 + \sin^2 \thg\, d\vf^2 \rh.
\label{aa.1}
\label{G90}
\ee
Considering geodesics in the equatorial plane $\thg = \pi/2$, there are two constants of motion representing the
energy and angular momentum per unit of mass:
\be 
\ve = \lh 1 - \frac{2M}{r} \rh \frac{dt}{d\tau}, \hs{2} \ell = r^2 \frac{d\vf}{d\tau}.
\label{aa.2}
\ee
Using eq.\ (\ref{aa.1}) the radial motion is then described by the equations
\be
\lh \frac{dr}{d\tau} \rh^2 = \ve^2 - \lh 1 - \frac{2M}{r} \rh \lh 1 + \frac{\ell^2}{r^2} \rh, \hs{1}
\frac{d^2r}{d\tau^2} = - \frac{M}{r^2} + \frac{\ell^2}{r^3} \lh 1 - \frac{6M}{r} \rh.
\label{aa.3}
\ee
It is not possible to solve the general equation for $r(\tau)$, except in special cases like circular orbits. However, 
one can exchange the proper time dependence for azimuth angular dependence by using the second equation 
(\ref{aa.2}), and determine $r(\vf)$ from
\be
\ell^2 \lh \frac{d}{d\vf} \frac{1}{r} \rh^2 = \ve^2 - \lh 1 - \frac{2M}{r} \rh \lh 1 + \frac{\ell^2}{r^2} \rh.
\label{aa.4}
\ee
This equation has solutions in terms of elliptic integrals. On the other hand, the geodesic deviation method allows to 
find approximate analytic solution for $r(\tau)$ itself starting from a special solvable geodesic, such as the circular 
orbit. 

Even though it does not produce explicit solutions for $r(\tau)$, the shape equation (\ref{aa.4}) contains useful 
information. To extract this information, we introduce a new variable $y$ \ct{jwvh_vu} such that
\be
r = \frac{a}{1 + e \cos y}, 
\label{aa.5}
\ee
where the parameters $(e, a)$ are implicitly given by
\be
\ba{l}
\dsp{ \ve^2 = \lh 1 - \frac{2M}{a} \rh \lh 1 + \frac{\ell^2}{a^2} \rh + \frac{e^2 \ell^2}{a^2} \lh 1 - \frac{6M}{a} \rh, }\\
 \\
\dsp{ a^2 - \frac{a \ell^2}{M} + \ell^2 \lh 3 + e^2 \rh = 0. }
\ea
\label{aa.6}
\ee
In the newtonian approximation $e$ represents the eccentricity of the elliptic orbit, and $(1 - e^2) a$ is the length
of the semi-major axis. Eq.\ (\ref{aa.4}) implies the full solution to satisfy 
\be
\lh \frac{dy}{d\vf} \rh^2 = 1 - \frac{2M}{a} \lh 3 + e \cos y \rh.
\label{aa.7}
\ee
Clearly, for large $a \gg M$ the second term is negligeable and we can take $y = \vf - \vf_0$, which returns the exact
newtonian orbit. Close to the horizon this is no longer applicable, as is shown for example by the existence of an
innermost stable circular orbit (ISCO) with $r = 6M$. 

The periastra of the geodesics (\ref{aa.5}) are reached for $y = 2 \pi n$. Eq.\ (\ref{aa.7}) shows, that the period in $y$
is not identical to the period in $\vf$, explaining the periastron shift in Schwarzschild-Droste co-ordinates. We can now 
derive integral expressions for this periastron shift in azimuth angle $\vf$ and in observer time $t$, as follows.
The total change $\Del \vf$ between two periastra is 
\be
\Del \vf = 2\pi + \del \vf =  \int_0^{2\pi} dy\, \frac{1}{\sqrt{ 1 - \frac{2M}{a} \lh 3+ e \cos y \rh }},
\label{aa.8}
\ee
where $\del \vf$ is the advance of the periastron compared to the previous one. Now using the conservation laws
we can also write an expression for the time lapse between two periastra
\be
\Del t = \frac{\ve a^2}{\ell}\, \int_0^{2\pi} dy\, \frac{1}{\lh 1 + e \cos y \rh^2 \lh 1 - \frac{2M}{a} \lh 1 + e \cos y \rh \rh
 \sqrt{ 1 - \frac{2M}{a} \lh 3 + e \cos y \rh }}.
\label{aa.9}
\ee
It follows plainly that the angular shift $\Del \vf$ and the time lapse $\Del t$ between two successive periastra are 
{\em independent} orbital characteristics, determined by the two independent parameters $(\ve, \ell)$ or equivalently 
$(a, e)$. 

The counting of parameters determining geodesics is simple: once geodesics are restricted to the equatorial plane
$\thg = \pi/2$, there remain three ordinary second-order differential equations for the co-ordinates $(t, r, \vf)$. The 
solutions dependend a priori on six constants of integration: three for the co-ordinates and three for the velocities.
However, the velocities are restricted by the normalization of the four-velocity $u^2 = -1$, cf.\ the first eq.\ (\ref{aa.3}). 
Therefore in practice we fix the origin of time and azimuth angle, the initial value of radial co-ordinate --which we
usually take to be the radial position of the periastron-- and two velocity parameters represented by the constants 
of motion $(\ve, \ell)$. 

\section{Driving terms for the second-order deviations \label{sab}}

In this section we present the explicit expressions for the coefficients $(a^{\mu}, b^{\mu}, c^{\mu})$ in the expansion
of the driving terms $\Sg^{\mu}$ for the second order deviations, eq.\ (\ref{5.4}), in the case of Schwarzschild geometry. 
The general expression for these terms is 
\be
\Sg^{\mu} = S^{\mu} + \frac{\og_1}{\og}\, T^{\mu},
\label{ab.1}
\ee
with $S^{\mu}$ as given in (\ref{3.1}) and (\ref{4.0}), after substitution $\tau \rightarrow \lb$, and $T^{\mu}$ defined by 
\be
T^{\mu} = - 4\, \frac{d^2 n^{\mu}}{d\lb^2} - 4 \brGam_{\lb\nu}^{\;\;\;\mu}\, u^{\lb} \frac{dn^{\nu}}{d\lb}.
\label{ab.2}
\ee
In the expansion of $S^{\mu}$ as defined in eq.\ (\ref{4.0}):
\[
S^{\mu}[n] = A_c^{\mu} \cos 2 \og \tau + A_s^{\mu} \sin 2 \og \tau + B_c^{\mu} \cos \og \tau + B_s^{\mu} \sin \og \tau + C^{\mu},
\]
the coefficients $A^t_c = A^r_s = A^{\vf}_c = 0$ vanish, whilst
\be
\ba{lll}
A^t _s & = & \dsp{ a^t\, =\, - \frac{2\og \ag}{R}\, \frac{1}{1 - \frac{2M}{R}} \lh n^r_c \rh^2, }\\
 & & \\
A^r_c & = & \dsp{ a^r\, =\, - \frac{M}{R^4} \frac{3 - \frac{M}{R} - \frac{18 M^2}{R^2}}{\lh 1 - \frac{2M}{R} \rh\lh 1 - \frac{3M}{R} \rh}
 \lh n^r_c \rh^2, }\\
 & & \\
A^{\vf}_s & = & \dsp{ a^{\vf}\, =\, - \frac{3\og \eta}{R} \lh n^r_c \rh^2. } 
\ea
\label{ab.3}
\ee
Next, $B^t_c = B^r_s = B^{\vf}_c = 0$, and
\be
\ba{lll}
B^t_s & = & \dsp{ - \frac{\og\ag}{R} \frac{1}{\lh 1 - \frac{2M}{R} \rh \lh 1 - \frac{3M}{R} \rh} \lh 4 - \frac{13M}{R} + \frac{6M^2}{R^2} \rh
 \Del_n n^r_c, }\\
 & & \\
B^r_c & = & \dsp{ - \frac{2M}{R^4} \frac{1}{\lh 1 - \frac{2M}{R} \rh \lh 1 - \frac{3M}{R} \rh^2} \lh 7 - \frac{36M}{R} + \frac{48M^2}{R^2} \rh 
 \Del_n n^r_c, }\\
 & & \\
B^{\vf}_s & = & \dsp{- \frac{\og\eta}{R} \frac{1}{1 - \frac{3M}{R}} \lh 5 - \frac{12M}{R} \rh  \Del_n n^r_c. }
\ea
\label{ab.4}
\ee
In addition there is a constant term in the $r$-component:
\be
\ba{lll}
C^r & = &\dsp{ - \frac{3M}{R^4} \frac{1 + \frac{M}{R}}{1 - \frac{3M}{R}}\, (n_c^r)^2  
 - \frac{3M}{2R^4} \frac{5 - \frac{34M}{R} + \frac{75M^2}{R^2} - \frac{54 M^3}{R^3}}{\lh 1 - \frac{3M}{R} \rh^3}\, \lh \Del_n \rh^2. }
\ea
\label{ab.4.1}
\ee
The Lindstedt-Poincar\'{e} correction terms (\ref{ab.2}) are of the form
\be
T^{\mu} = D^{\mu}_s \sin \og \lb + D^{\mu}_c \cos \og \lb + E^{\mu},
\label{ab.5}
\ee
with $D_c^t = D_s^r = D_c^{\vf} = 0$ and
\be
\ba{lll}
D_s^t & = & -2 \og \ag n_c^r, \\
 & & \\
D_c^r & = & \dsp{ - \frac{4M}{R^3} \frac{1}{1 - \frac{3M}{R}}\, n_c^r, }\\
 & & \\
D_s^{\vf} & = & - 2 \og \eta n_c^r, 
\ea
\label{ab.6}
\ee
whilst $E^t = E^{\vf} = 0$ and
\be
E^r = - 2 \kg\, \Del_n.
\label{ab.7}
\ee
Combining the results to compute
\[
b^t = B_s^t + \frac{\og_1}{\og}\, D_s^t, \hs{1} b^r = B^r_c + \frac{\og_1}{\og}\, D^r_c, \hs{1}
b^{\vf} = B^{\vf}_s + \frac{\og_1}{\og}\, D^{\vf}_s, 
\]
we find the coefficients $b^{\mu}$ to be of the form (\ref{5.11}) with
\be
\ba{lll}
F^t & = & \dsp{- \og \ag\, \frac{4 - \frac{13M}{R} + \frac{6M^2}{R^2}}{\lh 1 - \frac{2M}{R} \rh \lh 1 - \frac{3M}{R} \rh}, }\\
 & & \\
F^r & = & \dsp{ - \frac{2M}{R^3} \frac{7 - \frac{36M}{R} + \frac{48M^2}{R^2}}{\lh 1 - \frac{2M}{R} \rh \lh 1 - \frac{3M}{R} \rh^2}, }\\
 & & \\
F^{\vf} & = & \dsp{ - \og \eta\, \frac{5 - \frac{12M}{R}}{1 - \frac{3M}{R}}, }
\ea
\label{ab.8}
\ee
and
\be
\ba{lll}
G^t & = & \dsp{ 2 \og \ag, }\\
 & & \\
G^r & = & \dsp{ \frac{4M}{R^3} \frac{1}{1 - \frac{3M}{R}}, }\\
 & & \\
G^{\vf} & = & 2 \og \eta. 
\ea
\label{ab.9}
\ee
It then follows by eq.\ (\ref{5.12}), that
\be
\frac{\og_1}{\og} = - \frac{3}{2} \frac{1 - \frac{10M}{R} + \frac{18M^2}{R^2}}{\lh 1 - \frac{3M}{R} \rh \lh 1 - \frac{6M}{R} \rh}\, \frac{\Del_n}{R}. 
\label{ab.10}
\ee
Note that, whilst on the ISCO $R = 6M$ the fundamental frequency vanishes: $\og \rightarrow 0$, the Lindstedt-Poincar\'{e} frequency shift 
for finite $\Del_n$ is singular there: $\og_1 \rightarrow \infty$. 

Finally, the constant $c^r$ in the driving terms $\Sg^r$ is given by
\be
 c^r =  C^r + \frac{\og_1}{\og}\, E^r.
 \label{ab.12}
\ee
Using the value (\ref{ab.10}) for the ratio $\og_1/\og$, this leads to the result
\be
\ba{lll}
c^r & = & \dsp{ - \frac{3M}{R^4} \frac{1 + \frac{M}{R}}{1 - \frac{3M}{R}}\, (n_c^r)^2  
 + \frac{3M}{2 R^2} \lh \frac{\Del_n}{R} \rh^2 \frac{1 - \frac{26 M}{R} 
 + \frac{165 M^2}{R^2} - \frac{396M^3}{R^3} + \frac{324 M^4}{R^4}}{\lh 1 - \frac{3M}{R} \rh^3 \lh 1 - \frac{6M}{R} \rh}. }
\ea
\label{ab.11}
\ee

\np

\end{document}